\documentclass[twocolumn,showpacs,prl]{revtex4}
\usepackage[latin9]{inputenc}
\usepackage{color}
\usepackage{verbatim}
\usepackage{amsmath}
\usepackage{amssymb}
\usepackage{graphicx}
\usepackage{esint}
\usepackage{hyperref}

\makeatletter
\@ifundefined{textcolor}{}
{%
 \definecolor{BLACK}{gray}{0}
 \definecolor{WHITE}{gray}{1}
 \definecolor{RED}{rgb}{1,0,0}
 \definecolor{GREEN}{rgb}{0,1,0}
 \definecolor{BLUE}{rgb}{0,0,1}
 \definecolor{CYAN}{cmyk}{1,0,0,0}
 \definecolor{MAGENTA}{cmyk}{0,1,0,0}
 \definecolor{YELLOW}{cmyk}{0,0,1,0}
}

\usepackage{mathrsfs}

\draft

\makeatother

\begin{document}

\title{Symmetry Reduction and Boundary Modes for Fe-Chains on an s-wave Superconductor}

\author{Yu-Qin Chen, Yi-Ming Wu, and Xiong-Jun Liu\footnote{Corresponding author: xiongjunliu@pku.edu.cn}}
\affiliation{International Center for Quantum Materials and School of Physics, Peking University, Beijing 100871, China}
\affiliation{Collaborative Innovation Center of Quantum Matter, Beijing 100871, China}
\begin{abstract}
We investigate the superconducting phase diagram and boundary modes for a quasi-1D system formed by three Fe-Chains on an s-wave superconductor, motivated by the recent Princeton experiment. The $\vec l\cdot\vec s$ onsite spin-orbit term, inter-chain diagonal hopping couplings, and magnetic disorders in the Fe-chains are shown to be crucial for the superconducting phases, which can be topologically trivial or nontrivial in different parameter regimes. For the topological regime a single Majorana and multiple Andreew bound modes are obtained in the ends of the chain, while for the trivial phase only low-energy Andreev bound states survive. Nontrivial symmetry reduction mechanism induced by the $\vec l\cdot\vec s$ term, diagonal hopping couplings, and magnetic disorder is uncovered to interpret the present results. Our study also implies that the zero-bias peak observed in the recent experiment may or may not reflect the Majorana zero modes in the end of the Fe-chains.
\end{abstract}



\maketitle

\emph{Introduction}.-- The search for non-Abelian Majorana zero modes (MZMs)~\cite{Wilczek,Alicea2012,Franz}, which have potential applications to
fault-tolerant topological quantum computation \cite{Nayak,Ivanov,Sankar}, is a focus of research in condensed matter physics.
MZM exists in the vortex core of a two-dimensional (2D) $\left(p+ip\right)$-wave topological superconductor
(SC) \cite{Read}, and at the end of a 1D $p$-wave
SC \cite{Kitaev1}. Theoretical proposals showed that topological superconductivity can be obtained with heterostructures formed by conventional $s$-wave SC and topological insulators
\cite{Fu1} or semiconductors with a Zeeman splitting \cite{Sau0,Alicea,Roman0,Roman,Oreg,Potter}. In such devices, the spin-orbit (SO) interaction drives the original
$s$-wave SC into an effective $p$-wave SC, leading to MZMs when the system is in topologically nontrivial regime. Motivated by these proposals, recent experiments using semiconducting nanowire/$s$-wave SC heterostructures observed the zero-bias peak (ZBP) in the differential tunneling conductance
spectra \cite{Kouwenhoven,Deng,Das}, which is a suggestive signature of MZMs~\cite{Law,Flensberg,Wimmer,Liu}, while different theoretical interpretations are also available for the ZBP observed in the experiments \cite{Patrick2012PRL,Lee2012PRL}.

Very recently, a spatially resolved ZBP is observed by scanning tunneling spectroscopy (STM) in the chains of Fe atoms which exhibit ferromagnetic ordering and are placed on the surface of an $s$-wave SC (Pb)~\cite{Yazdani}. This study is motivated by but different from the earlier proposals of realizing 1D topological SC by adatoms with helical spin configurations on an $s$-wave SC~\cite{Spiral1}. It has been interpreted in theory that such Fe-chains may exhibit topological superconductivity since the Zeeman splitting, $s$-wave SC order, and Rashba SO interaction can be induced through the couplings between Fe atoms and Pb SC substrate~\cite{Yazdani,MacDonald2014}. Nevertheless, the valence electrons of Fe atoms occupy the $d$-orbital states which bring about ten bands for a single Fe-chain. Thus for the quasi-1D system formed by three Fe-Chains, as considered in the experiment, in general there are large number of bands crossing the Fermi energy, which may result in complicated phase diagram and boundary modes, besides the possible MZMs suggested in the experiment.

\begin{figure}[t]
\includegraphics[width=0.9\columnwidth]{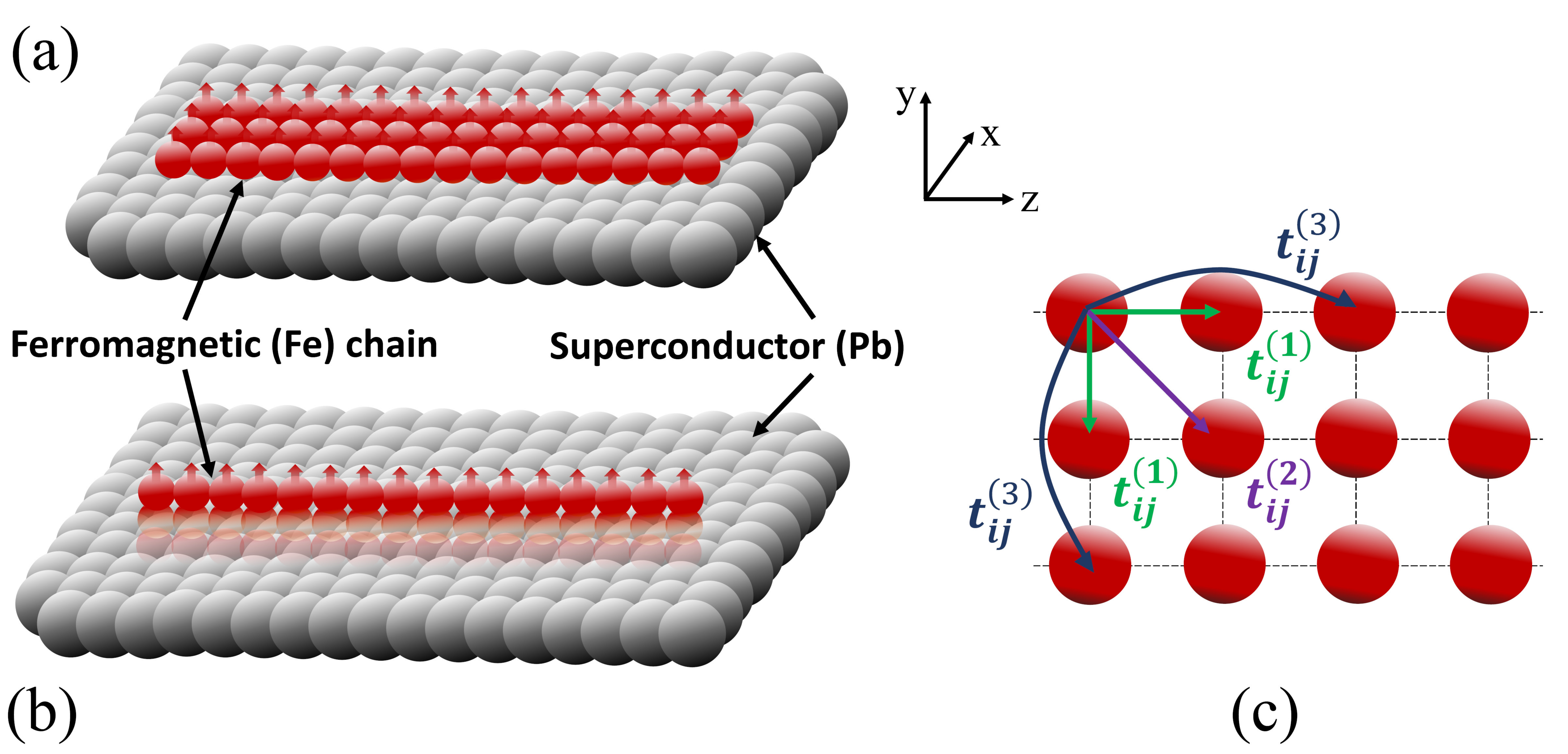} \caption{(Color online) Ferromagnetic Fe atomic chains stacked on the Pb substrate SC along $x$ (a) and $y$ (b) directions. The later configuration may be more relevant for the recent experiment~\cite{Yazdani}; (c) The hopping couplings between nearest-neighbor, next-nearest-neighbor, and next-next-nearest-neighbor sites ($t^{(l)}_{ij}, l=1,2,3$) are considered.}
\label{sketch}
\end{figure}

In this letter, we investigate the superconducting phase diagram and boundary modes for the three Fe-Chains placed on the Pb s-wave SC [Fig.~\ref{sketch} (a-c)]. We find that the $\vec l\cdot\vec s$ term for $d$-orbital electrons, inter-chain diagonal hopping couplings, and magnetic disorders in the Fe-chains play crucial roles in determining the symmetry classes of the superconducting phases. For the topological phase we show that a single MZM and multiple Andreev bound states (ABSs) are obtained in the ends of the chain, while in the trivial regime only low-energy ABSs exist. The novel symmetry reduction mechanism is uncovered clearly to interpret these results.

\emph{Model}.-- In general the configuration of the Fe atomic chains on the Pb surface may be complicated due to the strong Fe-Pb bonding. To capture the essential physics, we consider here the simplest situation that the three Fe chains sit in parallel on the surface and along $z$ direction, with intra-chain and inter-chain hopping couplings being taken into account. In particular, without loss of generality, we consider here the hopping couplings between up to next-next-nearest neighbor sites [Fig.~\ref{sketch} (c)]. Moreover, Two different configurations with chains stacked along $x$ and $y$ directions [Fig.~\ref{sketch} (a,b)] are considered. With the Slater-Koster basis~\cite{SK} the effective model without disorder can be described by
\begin{eqnarray}\label{eqn:H1}
H_{\rm triple}&=&-\mu_F\sum_{\alpha}\sum_{\bold r,\sigma}c^\dag_{\alpha \sigma}(\bold r)c_{\alpha \sigma}(\bold r)+\nonumber\\
&&+\sum_{\alpha,\alpha'}\sum_{j,\sigma}\sum_{\bold r\neq\bold r'}t^{(j)}_{\alpha\alpha'}(\bold r,\bold r')c^\dag_{\alpha \sigma}(\bold r)c_{\alpha' \sigma}(\bold r')+h.c.\nonumber\\
&&+\sum_{\alpha}\sum_{\bold r,\sigma\sigma'}c^\dag_{\alpha \sigma}(\bold r)Js^y_{\sigma\sigma'}c_{\alpha \sigma'}(\bold r)\nonumber\\
&&+i\lambda_{\rm so}\sum_{\alpha,\alpha'}\sum_{\bold r,\sigma,\sigma'}c^\dag_{\alpha \sigma}(\bold r)(\vec l_{\alpha\alpha'}\cdot\vec s_{\sigma\sigma'})c_{\alpha \sigma'}(\bold r)\nonumber\\
&&+it_{R}\sum_{\alpha}\sum_{\bold r,\sigma,\sigma'}c^\dag_{\alpha \sigma}(\bold r)s^x_{\sigma\sigma'} c_{\alpha \sigma'}(\bold r+a\hat e_z)+h.c.\nonumber\\
&&+\sum_{\alpha,\bold r}\left[\Delta_s(\bold r)c_{\alpha\uparrow}(\bold r)c_{\alpha\downarrow}(\bold r)+h.c.\right],
\end{eqnarray}
where $c_{\alpha\sigma}$ ($c^\dag_{\alpha\sigma}$) is the annihilation (creation) operator for the $d$-orbital electron, with $\alpha$ and $\sigma=\uparrow,\downarrow$ labeling the five orbital states and spin states, respectively, $\mu_F$ is the chemical potential, $J$ is the Stoner-theory spin splitting energy with magnetization along $y$ direction, $\lambda_{\rm so}$ represents the onsite SO coefficient, $t_R$ is the Rashba SO coefficient, and $\Delta_s$ is the proximity induced $s$-wave SC order in the Fe chains. The coefficients $t^{(j)}_{\alpha\alpha'}(\bold r,\bold r')$ represent the nearest-neighbor (for $j=1$), next-nearest-neighbor or diagonal (for $j=2$), and next-next-nearest-neighbor (for $j=3$) hopping couplings in the Slater-Koster approximation [Fig.~\ref{sketch} (b)]. The numerical magnitudes of the parameters are given in the Supplementary Material~\cite{SI} according to the density function theory calculation in Ref.~\cite{Yazdani}. The spin splitting $J\sim2.7$eV is the largest energy scale in the above formula, which leads to the full spin polarized bands in the Fermi energy. In the presence of the Rashba SO interaction induced by the interface hybridization between Fe $d$-orbital and Pb $p$-orbital states, the proximity induced $s$-wave SC may be driven into an effective $p$-wave topological SC, which can support MZMs at the Fe-chain ends~\cite{Roman0,Roman,Oreg}.

From the Slater-Koster basis, one can check that the onsite $\vec l\cdot\vec s$ and inter-chain hopping ($t^{(2)}_{\alpha\alpha'}$) terms can induce the couplings between different $d$-orbital states. These couplings can qualitatively affect the symmetries respected by the Hamiltonian. We shall show that the these coupling terms and the magnetic disorder, which shall be considered later, play the crucial roles in determining the symmetry classes and thus the superconducting phases. Without such terms, the system generically has multiple MZMs localized in each end of Fe-chains, while the presence of such couplings can mix the MZMs, giving rise to low-energy ABSs. For the sake of a clear understanding of these effects, in the following we examine the present Fe-chain system step by step.

\begin{figure}[t]
\includegraphics[width=1\columnwidth]{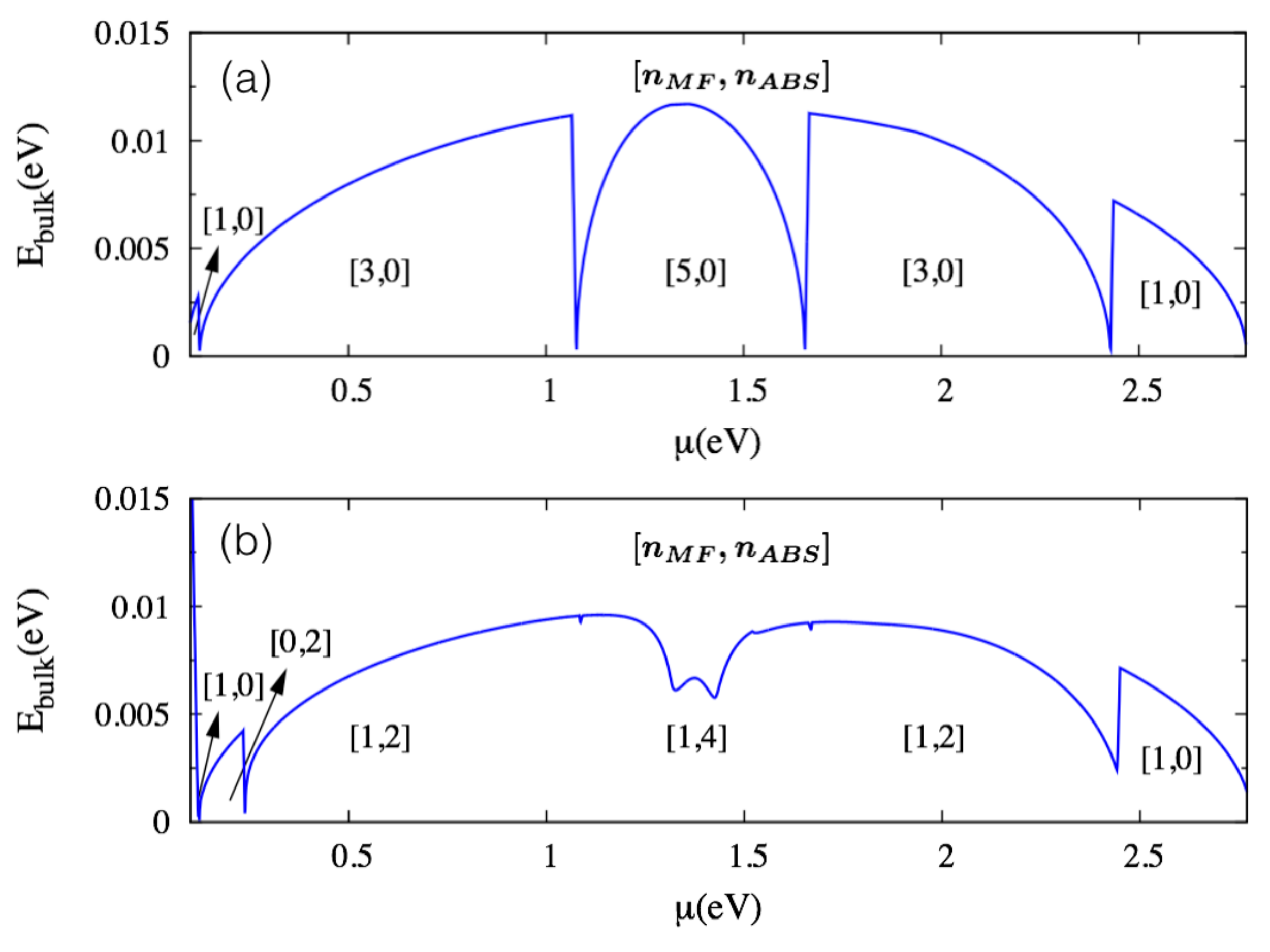} \caption{(Color online) Phase diagram and boundary modes for a single Fe chain with (a) $\lambda_{\rm so}=0$ and (b) $\lambda_{\rm so}=0.15$eV for the onsite $\vec l\cdot\vec s$ term. Other parameters are taken as $t_R=0.1$eV, $\Delta_s=0.04$eV, and the remaining hopping coefficients are given in Supplementary Material~\cite{SI}. The trivial phase corresponds to $n_{\rm MF}=0$ in the notation $[n_{\rm MF},n_{\rm ABSs}]$.}
\label{phase1}
\end{figure}

\emph{Single Fe-chain case}.-- Let us first consider the simplest situation with a single Fe Chain. This is equivalent to study the Hamiltonian~\eqref{eqn:H1} without inter-chain couplings. The single chain Hamiltonian in the momentum space reads ${\cal H}_{\rm single}(\lambda_{\rm so},k_z)=(2V_1\cos k_z+2V_3\cos 3k_z-\mu_F)\tau_z+2t_R\sin k_z s_x-J/2 s_y+\Delta_ss_y\tau_y+\lambda_{\rm so}\vec l\cdot\vec s\tau_z$, where $V_1$ and $V_2$ are matrices corresponding to hopping terms $t_{ij}^{(1)}$ and $t_{ij}^{(3)}$, respectively~\cite{SI}. A key feature of the present system is that the symmetry class of the Hamiltonian depends on onsite SO term. In the absence of $\vec l\cdot\vec s$ term, namely, if $\lambda_{\rm so}=0$, we find that the Hamiltonian respects both the time-reversal (TR) symmetry $T$ and charge conjugation symmetry $\cal C$ defined via $T{\cal H}_{\rm single}(\lambda_{\rm so}=0,k_z)T^{-1}={\cal H}_{\rm single}(\lambda_{\rm so}=0,-k_z)$, and ${\cal C}{\cal H}_{\rm single}(\lambda_{\rm so}=0,k_z){\cal C}^{-1}=-{\cal H}^*_{\rm single}(\lambda_{\rm so}=0,-k_z)$, with
\begin{eqnarray}\label{eqn:symmetry1}
T=Ks_z\tau_z, \ {\cal C}=\tau_x, \ T^2={\cal C}^2=1.
\end{eqnarray}
Here $K$ is the complex conjugate operator. The above result implies that ${\cal H}_{\rm single}(\lambda_{\rm so}=0,k_z)$ belongs to the BDI symmetry class according to the ten-fold topological classification~\cite{A-Z,Ryu,Kitaev2009,Wen}, which can protect integer number of MZMs at each end~\cite{Tewari2012,Chakravarty2012,Beenakker2012PRB,Vic2014,Sarma2015} [see Fig.~\ref{phase1} (a)]. Note that the symmetries $T,{\cal C}$ do not transform orbital states, implying that each orbital band at the Fermi energy contributes one MZM. Thus the multiple MZMs correspond to the multiple orbital subbands crossing the Fermi energy.

Once the $\vec l\cdot\vec s$ term is switched on, the TR symmetry $T$ defined in Eq.~\eqref{eqn:symmetry1} is broken. Indeed the multiple $d$-orbital states can be equivalently treated as the degree of freedom of a ``synthetic" transverse dimension. Then the $\vec l\cdot\vec s$ and 1D Rashba terms render the SO couplings in the synthetic and physical dimensions, respectively, giving an effective 2D SO coupling which cannot be real and thus breaks the above $T$ symmetry. Only the charge conjugation symmetry keeps and the symmetry class of the system is reduced from BDI class to D class, with the topology being classified by a $Z_2$ invariant, calculated by $(-1)^{\nu}={\rm sgn}\{{\rm Pf}[{\cal H}(k_z=0)\tau_x]{\rm Pf}[{\cal H}(k_z=\pi)\tau_x]\}$. The topologically nontrivial (trivial) phase corresponds to $\nu=1 (0)$. From Fig.~\ref{phase1} (b) we can see that while in the most region the phase is topological, there are small regions which are topologically trivial. In the topological phase, only a single MZM is obtained in each end of the chain, with several ABSs coexisting. These ABSs originate from the mixing between MZMs obtained in Fig.~\ref{phase1} (a) by the $\vec l\cdot\vec s$ symmetry-breaking term. In the trivial phase, only low-energy ABSs are obtained.

\begin{figure}[t]
\includegraphics[width=1\columnwidth]{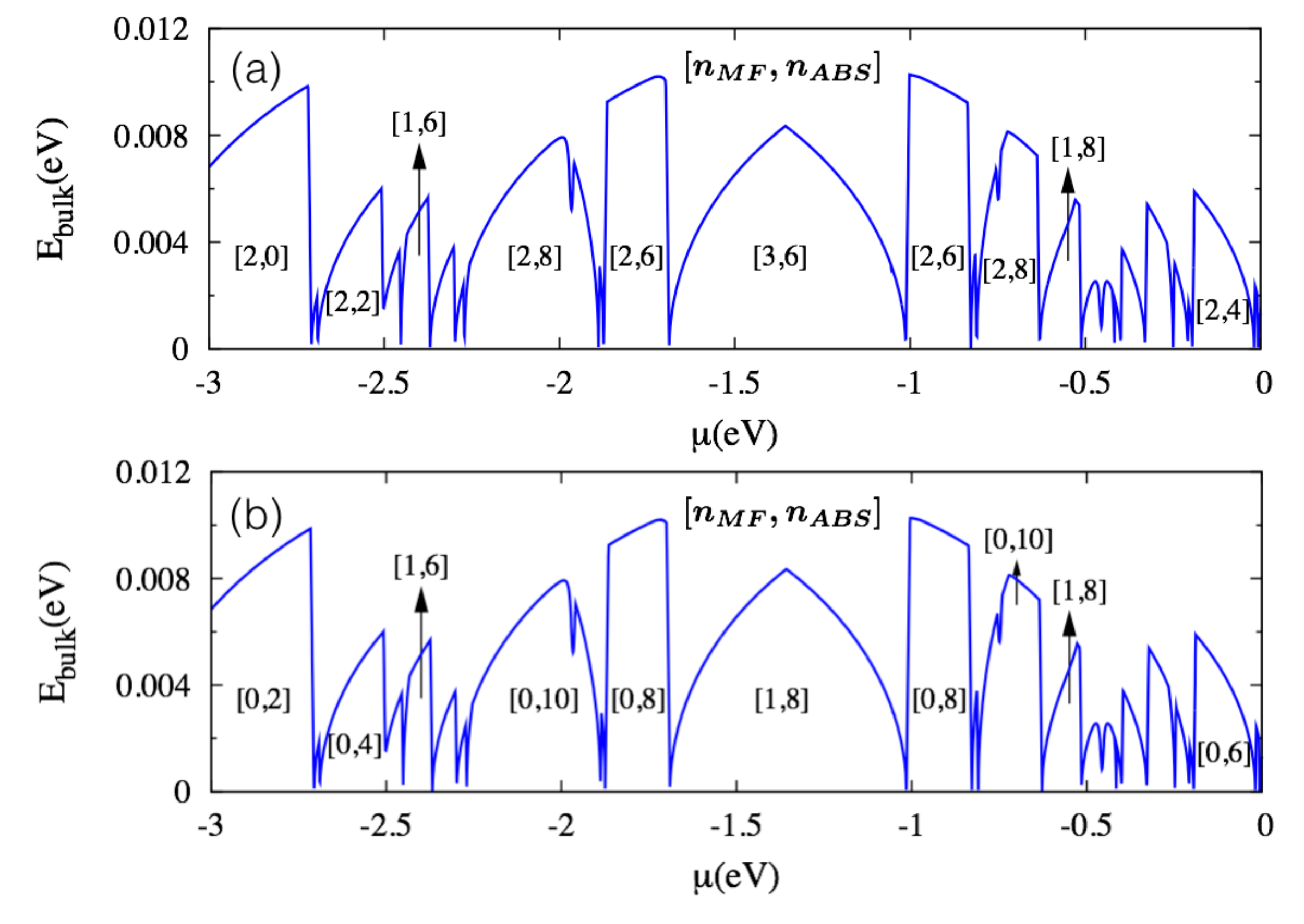} \caption{(Color online) Phase diagrams and boundary modes for the triple Fe-chain case with (a) $t^{(2)}_{\alpha\alpha'}=0$ and (b) $t^{(2)}_{\alpha\alpha'}=0.01t^{(1)}_{\alpha\alpha'}$ (and disorder amplitude $\delta J^x_{\rm max}=10$meV for the configuration (b) of Fig.~\ref{sketch}). Other parameters are $\lambda_{\rm so}=60$meV, $t_R=0.1$eV, $\Delta_s=0.04$eV, and the remaining hopping coefficients are given in Supplementary Material~\cite{SI}.}
\label{phase2}
\end{figure}

\emph{Triple Fe-chain case}.-- Now we turn to the three Fe-chain model given in Eq.~\eqref{eqn:H1}. We first consider the configuration (a). It will be shown that in this case the diagonal hopping coupling term also becomes crucial in determining the phases. To see this effect clearly, we parameterize the three-chain Bloch Hamiltonian ${\cal H}_{\rm triple}(t^{(2)}_{\alpha\alpha'},k_z)$ as function of $t^{(2)}_{\alpha\alpha'}$. It is interesting that a new set of TR ($\tilde T$) and charge conjugation ($\tilde{\cal C}$) symmetries are found when $t^{(2)}_{\alpha\alpha'}=0$, satisfying $\tilde T{\cal H}_{\rm triple}(t^{(2)}_{\alpha\alpha'}=0,k_z)\tilde T^{-1}={\cal H}_{\rm triple}(t^{(2)}_{\alpha\alpha'}=0,-k_z)$, and $\tilde{\cal C}{\cal H}_{\rm triple}(t^{(2)}_{\alpha\alpha'}=0,k_z)\tilde{\cal C}^{-1}=-{\cal H}^*_{\rm triple}(t^{(2)}_{\alpha\alpha'}=0,-k_z)$. Here $\tilde{\cal C}=\cal C$, while the TR symmetry is given by
\begin{eqnarray}\label{eqn:symmetry2}
\tilde T=UU^TK s_z\tau_z,\
U=\frac{1}{\sqrt{2}}{\left[
\begin{matrix}
i  & 0 & 0 & 0 & -i\\
0  & i & 0 & -i & 0\\
0  & 0 & \sqrt{2} & 0 & 0\\
0  & 1 & 0 & 1 & 0\\
1  & 0 & 0 & 0 & 1\\
\end{matrix} \right]}.
\end{eqnarray}
Here $U$ is a local unitary matrix acting on the five $d$-orbital bases $[d_{xy},d_{xz},d_{z^2},d_{yz},d_{x^2-y^2}]^T$, and one can verify that $\tilde T^2=1$. 
Thus the Hamiltonian ${\cal H}_{\rm triple}(t^{(2)}_{\alpha\alpha'}=0)$ belongs to a new BDI symmetry class characterized by $\tilde T$ and $\tilde{\cal C}$, and can support multiple MZMs. On the other hand, since the onsite $\vec l\cdot\vec s$ term breaks the TR symmetry $T$ as defined in the single-chain model and leads to low-energy ABSs, in general there are both multiple MZMs and multiple ABSs in the present three-chain system, as shown numerically in Fig.~\ref{phase2} (a). The ABSs are due to the couplings in MZMs induced by the $\vec l\cdot\vec s$ SO term.

Similar as the result in the single-chain system, the diagonal hopping in configuration (a) can break the new TR symmetry of ${\cal H}_{\rm triple}(t^{(2)}_{\alpha\alpha'}=0,k_z)$ by verifying that $\tilde T{\cal H}_{\rm triple}(t^{(2)}_{\alpha\alpha'},k_z)\tilde T^{-1}\neq{\cal H}_{\rm triple}(t^{(2)}_{\alpha\alpha'},-k_z)$ for $t^{(2)}_{\alpha\alpha'}\neq0$. This implies that the symmetry class of the system is again reduced from the new BDI class to the D class, and the topology is classified by the $Z_2$ invariant. Therefore, the diagonal hopping terms can couple the remaining multiple MZMs, with only a single MZM surviving if the total number of MZMs is odd for the regime without diagonal hopping couplings. On the contrary, if the number of MZMs for ${\cal H}_{\rm triple}(t^{(2)}_{\alpha\alpha'}=0,k_z)$ is even, the diagonal hopping term drives the system into a trivial phase with only low-energy ABSs existing. We plot the phase diagram by numerical calculation in Fig.~\ref{phase2} (b). It is clear that both topologically nontrivial and trivial phases are obtained in the large ranges of chemical potential, with each gap closing point separating a topological phase and a trivial phase. Due to multiple subbands crossing the Fermi energy, there are multiple ABSs obtained in almost all the different parameter regimes.

We emphasize that for triple-chain case the symmetry-breaking mechanism also requires the magnetization to be perpendicular to the Fe-chain stacking direction~\cite{SI}. This implies that for the second configuration [Fig.~\ref{sketch} (b)], even the diagonal hopping term cannot break the $\tilde T$ symmetry since the averaging magnetization is parallel to the stacking $(y)$ direction. Nevertheless, in the realistic system, this symmetry can be broken when random magnetic disorder with magnetization $\delta\vec J(\bold r)=\delta J^x\hat e_x$ along $x$ direction is present, giving the disorder Hamiltonian $V_{\rm dis}=\sum_{\alpha}\sum_{\bold r,\sigma\sigma'}c^\dag_{\alpha \sigma}(\bold r)\delta J^x(\bold r)s^x_{\sigma\sigma'}] c_{\alpha \sigma'}(\bold r)$. The further inclusion of the magnetic disorder $V_{\rm dis}$ in the configuration of Fig.~\ref{sketch} (b) leads to the same phase diagram shown in Fig.~\ref{phase2} (b).

\emph{Minigap and tunneling spectra}.-- With the existence of multiple ABSs in the end, it is important to calculate the minigap, defined as the energy of the lowest ABSs $E_{\rm mini}=\min\{E_{\rm ABS}\}$, of the real system. A sizable minigap is necessary to distinguish the topologically nontrivial phase from the trivial phase by STM measurement. The eigenvalues of the boundary modes at the end can be calculated with surface Green's function through iteration methods~\cite{surface}. The calculation can be performed by transfer matrix method. For the present 1D system with next-next-nearest-neighbor hopping couplings, one can separate the Fe-Chains into many principle segments (PSs) along $z$ axis, with each PS containing $q\geq2$ Fe atoms along the chain direction (the total Fe atom number in a PS is then $3q$). For the case with magnetic disorder, one needs to take $3q\gg1$ to avoid the numerical error~\cite{SI}. The Green's function of the system is denoted by $G_{n,n'}^{l,l'}(\omega)$, where $n,n'$ are the PS indices and $l,l'$ ($=1,...,q$) denote the $q$ atomic layers in each PS. The surface Green's function corresponds to $n=n'=l=l'=0$, and can be solved through
\begin{eqnarray}\label{eqn:surface1}
G_{00}(\omega)=\frac{I}{\omega+i\delta^+-H_{00}-H_{01}T}.
\end{eqnarray}
Here $I$ is a unit matrix, $H_{00}$ is the block Hamiltonian of the surface PS, and $H_{01}$ represents the couplings between the surface PS and the next PS which include the hopping couplings and the Rashba SO term. The transfer matrix $T$ is obtained by iteration method~\cite{SI}
\begin{eqnarray}\label{eqn:surface2}
T(\omega)=t_0+\sum_{n=1}^{N_c}\prod_{m=0}^{n-1}t_n\tilde t_m,
\end{eqnarray}
where $t_0=(\omega-H_{00})^{-1}H_{01}^\dag, \tilde t_0=(\omega-H_{00})^{-1}H_{01}$, $a_m=(I-t_{m-1}\tilde t_{m-1}-\tilde t_{m-1}t_{m-1})^{-1}a_{m-1}^2$ with $a_m=t_m$ or $\tilde t_m$, and $N_c$ is a cut-off. The local density of states are then obtained by
\begin{eqnarray}\label{eqn:LDOS}
\rho(\omega;n,l)=-\frac{1}{\pi}\Im[G_{n,n}^{l,l}(\omega)], \ n=l=0,
\end{eqnarray}
with which one can determine the spectra of the boundary modes.
With the obtained surface Green's function one can further calculate the tunneling conductance by considering the tunneling coupling between a metallic lead and the end area of the chains.

\begin{figure}[t]
\includegraphics[width=1\columnwidth]{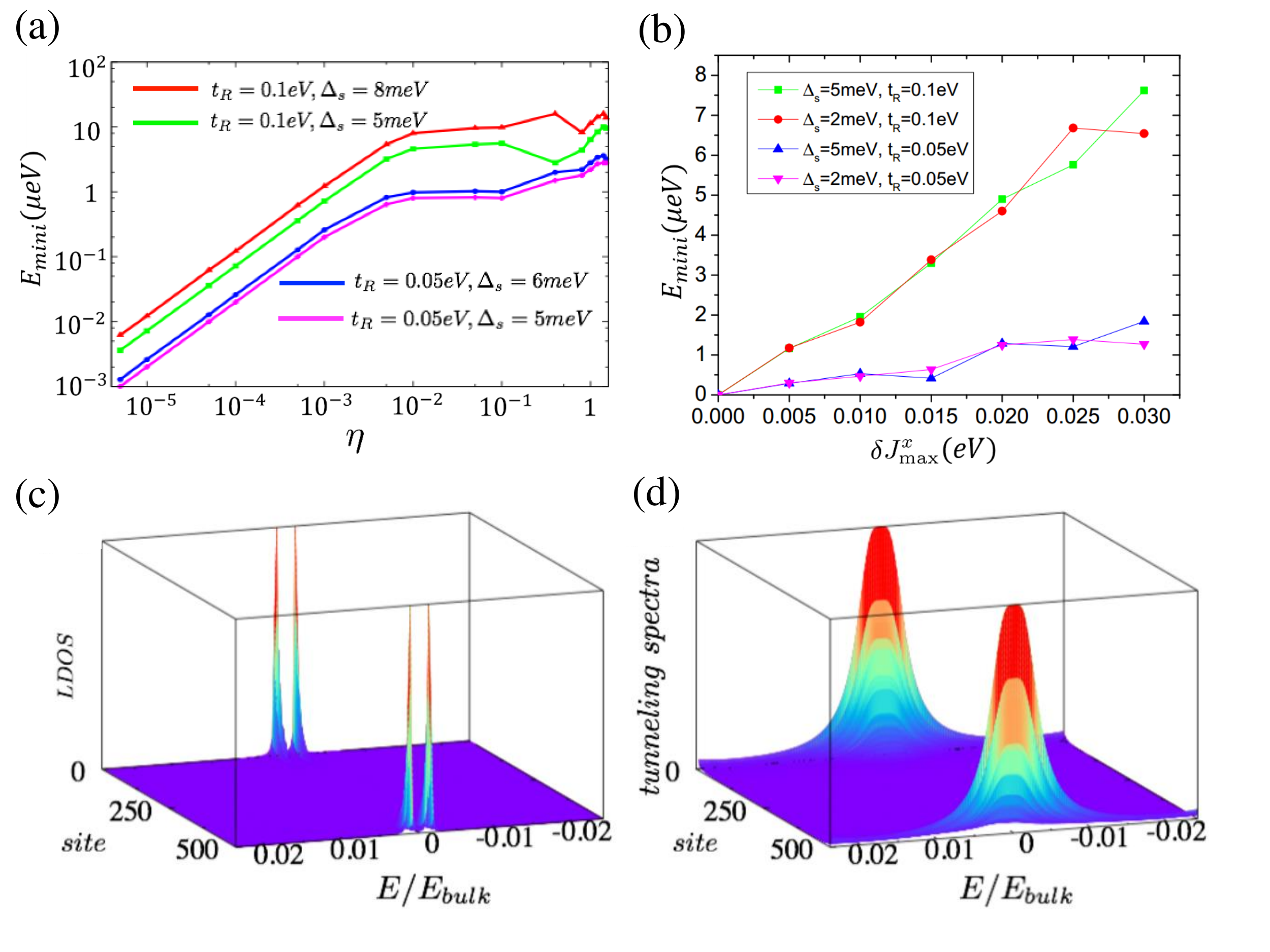} \caption{(Color online) Lowest ABSs for the trivial phase with $\lambda_{\rm so}=60$meV and $\mu=-1.0$eV. Minigap as a function of $\eta$ (a), and disorder amplitude $\delta J^x_{\rm max}$ with $\eta=1$ (b). (c) The local density of states (LDOS) contributed from the lowest ABSs for $\eta=1$, and (d) the spatially resolved tunneling spectra with a tunneling energy of $1.0E_{\rm mini}$.}
\label{Fig.4}
\end{figure}
The numerical results in the experimentally relevant parameter regimes are shown in Fig.~\ref{Fig.4}, where (a) and (b) shows the minigap as functions of the the ratio $\eta=4\sqrt{2}t_{\alpha\alpha'}^{(2)}/t^{(1)}_{\alpha\alpha'}$ between the diagonal hopping and nearest-neighbor hopping strengths with fixed $t^{(1)}_{\alpha\alpha'}$, and the amplitude of the random magnetic disorder $\delta J^x_{\rm max}$, respectively. The physical regime corresponds to $\eta=1$ under the Slater-Koster approximation~\cite{Harrison}. The coefficient of the onsite SO coupling $\vec l\cdot\vec s$ is fixed with $\lambda_{\rm so}=60$meV~\cite{Yazdani}, which may determine the minigap of the system if $\eta$ (and $\delta J^x_{\rm max}$) is large. Fig.~\ref{Fig.4} (a,b) shows that the minigap in a topologically trivial regime increases with the diagonal hopping coupling, and is typically a few $\mu$eV for $\eta\sim1$ and $\delta J^x_{\rm max}<30$meV. The local density of states are shown in Fig.~\ref{Fig.4} (c), which reflects that no MZMs but low-energy ABSs exist in the topologically trivial phase. However, from Fig.~\ref{Fig.4} (d) we can see that when the tunneling energy is over the minigap, the tunneling spectra manifests a ZBP, which brings about challenges to distinguish the topological phase from trivial phase by STM measurement. The more precise measurement is desired to identify the MZMs.

\emph{Discussion and conclusion}.-- Several issues are worthwhile to be mentioned. Firstly, the configurations we consider are straight and parallel Fe-chains within a single ($x-z$ or $y-z$) plane. In the real system the chains may be staggered or not in-plane due to complexity of the Fe-Pb interface. A staggered and not-in-plane configuration can generically bring about additional hopping couplings which also contribute to the symmetry reduction from the second type of BDI class to D class. Secondly, it was shown that the strong coupling between the Fe atoms and the substrate Pb atoms can strongly renormalize the Fermi velocities of the Fe bands~\cite{Oppen2015}. This effect is essentially because the transverse wave functions (penetrating into the substrate SC) of Fe electrons have strong dependence on the momentum $k_z$ due to the interface couplings, which greatly suppresses the dependence of energies on $k_z$. Nonetheless, if the interface couplings between Fe and Pb atoms do not break the TR symmetries $T$ and $\tilde T$ defined in the present work, the main results predicted here shall not be affected by the renormalization of Fermi velocities. A more detailed study of this effect will be performed in the next work. Finally,
The long-range hopping couplings along the chain and transverse direction do not affect the TR symmetries. Therefore, taking into account hopping couplings between Fe atoms with even larger distance does not change our results.

In conclusion, we have studied the superconducting phase diagram and boundary modes for a quasi-1D system formed by three Fe-Chains on an s-wave SC. We uncovered a nontrivial symmetry reduction mechanism with two different types of BDI classes (characterized by $Z$ invariant) reduced to D classes (with $Z_2$ invariant) by the onsite $\vec l\cdot\vec s$ term for $d$-orbital electrons, inter-chain diagonal hopping, and magnetic disorder couplings, which governs the properties of boundary modes in the topologically nontrivial or trivial phases. For the topological regime a single Majorana and multiple Andreew bound modes are obtained in each end of the chain, while for the trivial phase only low-energy Andreev bound states survive. Our results call for further experimental studies to identify the MZMs out of low-energy ABSs.

\begin{acknowledgments}.
\end{acknowledgments}
We thank Fa Wang, Xin Liu, and Ting-Fung Jeffrey Poon for helpful discussions. This work
is supported in part by the Thousand-Young-Talent Program of China.
 \bibliographystyle{apsrev}



\onecolumngrid

\renewcommand{\thesection}{S-\arabic{section}}
\renewcommand{\theequation}{S\arabic{equation}}
\setcounter{equation}{0}  
\renewcommand{\thefigure}{S\arabic{figure}}
\setcounter{figure}{0}  

\section*{Symmetry Reduction and Boundary Modes for Fe-Chains on an s-wave Superconductor --- Supplementary Material}


\subsection{Single Fe-chain case}

Let the spin be polarized along $\vec{y}$ direction, the whole single chain Hamiltonian in the momentum space can be written as ${\cal H}_{\rm single}(\lambda_{\rm so},k_z)={\cal H}_{\rm single}(\lambda_{\rm so}=0,k_z)+{\cal H}_{\rm so}$, where ${\cal H}_{\rm single}(\lambda_{\rm so}=0,k_z)=(2V_1\cos k_z+2V_3\cos3k_z-\mu_F)I_B\otimes \tau_z\otimes s_0+2t_R\sin k_zI_B\otimes \tau_0\otimes s_x-J/2I_B\otimes \tau_0\otimes  s_y+2\Delta_sI_B\otimes \tau_y\otimes s_y$, with $V_1$ and $V_3$ are matrices corresponding to hopping terms $t_{ij}^{(1)}$ and $t_{ij}^{(3)}$, respectively, given in Tablets I and II.
\begin{table}[!hbp]
\begin{tabular}{|c|c|c|c|c|c|}
\hline
$t^{(1)}_{ij}$ or\  $V_1$(eV) & $d_{xy}$ & $d_{xz}$ & $d_{z^2}$ & $d_{yz}$ & $d_{x^2-y^2}$ \\
\hline
$d_{xy}$ & -0.1445 & 0 & 0 & 0 & 0 \\
\hline
$d_{xz}$ & 0& 0.5760 & 0 & 0 & 0 \\
\hline
$d_{z^2}$ & 0 & 0 & -0.6702 & 0 & 0\\
\hline
 $d_{yz}$ & 0 & 0 & 0 & 0.5760 & 0 \\
\hline
$d_{x^2-y^2}$ & 0 & 0 & 0 & 0 & -0.1445 \\
\hline
\end{tabular}
\caption{Nearest-neighbor hopping couplings along $\vec{z}$ direction.}
\end{table}
\begin{table}[!hbp]
\begin{tabular}{|c|c|c|c|c|c|}
\hline
$2^5t^{(3)}_{ij}$ or $2^5V_3$(eV) & $d_{xy}$ & $d_{xz}$ & $d_{z^2}$ & $d_{yz}$ & $d_{x^2-y^2}$ \\
\hline
$d_{xy}$ & -0.1445 & 0 & 0 & 0 & 0 \\
\hline
$d_{xz}$ & 0& 0.5760 & 0 & 0 & 0 \\
\hline
$d_{z^2}$ & 0 & 0 & -0.6702 & 0 & 0\\
\hline
 $d_{yz}$ & 0 & 0 & 0 & 0.5760 & 0 \\
\hline
$d_{x^2-y^2}$ & 0 & 0 & 0 & 0 & -0.1445 \\
\hline
\end{tabular}
\caption{Next-next-nearest-neighbor hopping couplings along $\vec{z}$ direction~\cite{SIHarrison}.}
\end{table}
$I_B$ is the unit matrix of band degree, $\vec{\tau}$ and $\vec{s} $ represent Pauli matrices of particle-hole and spin respectively. The onsite spin-orbit (SO) coupled Hamiltonian reads
\begin{equation}
{\cal H}_{\rm so}=\frac{1}{2} \lambda_{\rm so}\cdot
\left(
  \begin{array}{cccccccccc}
    0 & 0 & -s_y &0 &0  & 0  & s_x  & 0  & -2is_z  & 0     \\
    0 & 0 & 0 & -s_y  & 0  & 0  & 0  & -s_x  & 0  & -2is_z   \\
    -s_y & 0 & 0 &0   & \sqrt{3}s_x  & 0  & is_z  & 0  & s_x  & 0  \\
    0 & -s_y & 0 &0   & 0  & -\sqrt{3}s_x  & 0  & is_z  & 0  & -s_x   \\
    0 & 0 & \sqrt{3}s_x &0   & 0  & 0  & -\sqrt{3}s_y  & 0  & 0  & 0   \\
    0 & 0 & 0 &-\sqrt{3}s_x  & 0  & 0  & 0  & -\sqrt{3}s_y  & 0  & 0   \\
    s_x & 0 & -is_z &0   & -\sqrt{3}s_y  & 0  & 0  & 0  & s_y  & 0   \\
    0 & -s_x & 0 &-is_z   & 0  & -\sqrt{3}s_y  & 0  & 0  & 0  & s_y   \\
    2is_z & 0 & s_x &0   & 0  & 0  & s_y  & 0  & 0  & 0  \\
    0 & 2is_z & 0 &-s_x   & 0  & 0  & 0  & s_y  & 0  & 0  \\
  \end{array}
\right),
\end{equation}
Note that ${\cal H}_{\rm so}$ is written in the Nambu space.

When there is no onsite SO interaction, the Hamiltonian  ${\cal H}_{\rm single}(\lambda_{\rm so}=0,k_z)$ respects both the time-reversal (TR) symmetry $T$ and charge conjugation symmetry $\cal C$ defined via $T{\cal H}_{\rm single}(\lambda_{\rm so}=0,k_z)T^{-1}={\cal H}_{\rm single}(\lambda_{\rm so}=0,-k_z)$, and ${\cal C}{\cal H}_{\rm single}(\lambda_{\rm so}=0,k_z){\cal C}^{-1}=-{\cal H}^*_{\rm single}(\lambda_{\rm so}=0,-k_z)$, with
\begin{eqnarray}
T=K I_B\otimes \tau_z \otimes s_z, \ {\cal C}=I_B\otimes \tau_x \otimes s_0, \ T^2={\cal C}^2=1.
\end{eqnarray}
Here $K$ is the complex conjugate operator. The above result implies that ${\cal H}_{\rm single}(\lambda_{\rm so}=0,k_z)$ belongs to the BDI symmetry class.

It is easy to check that $T{\cal H}_{\rm so}T^{-1}\neq{\cal H}_{\rm so}$, ${\cal C}{\cal H}_{\rm so}{\cal C}^{-1}=-{\cal H}^*_{\rm so}$, thus the TR symmetry $T$ defined in Eq.(2) is broken by the onsite $\vec l\cdot\vec s$ term. Only the charge conjugation symmetry keeps. As a result, when $\vec l\cdot\vec s$ term is present, the symmetry class of the system is reduced from BDI class to D class.

\subsection{Triple Fe-chain for the configuration (a)}

For the configuration (a) the spin is polarized along $\vec{y}$ direction, and the chains are stacked along $x$ direction. 
We have $\tilde T{\cal H}_{\rm triple}(t^{(2)}_{\alpha\alpha'}=0,k_z)\tilde T^{-1}={\cal H}_{\rm triple}(t^{(2)}_{\alpha\alpha'}=0,-k_z)$, and $\tilde{\cal C}{\cal H}_{\rm triple}(t^{(2)}_{\alpha\alpha'}=0,k_z)\tilde{\cal C}^{-1}=-{\cal H}^*_{\rm triple}(t^{(2)}_{\alpha\alpha'}=0,-k_z)$, where
\begin{eqnarray}\label{eqn:SIsymmetry2}
\tilde T=UKU^{\dag} s_z\tau_z,\
U=\frac{1}{\sqrt{2}}{\left[
\begin{matrix}
i  & 0 & 0 & 0 & -i\\
0  & i & 0 & -i & 0\\
0  & 0 & \sqrt{2} & 0 & 0\\
0  & 1 & 0 & 1 & 0\\
1  & 0 & 0 & 0 & 1\\
\end{matrix} \right]}.
\end{eqnarray}
Here $U$ is a local unitary matrix acting on the five SK $d$-orbital bases $[d_{xy},d_{xz},d_{z^2},d_{yz},d_{x^2-y^2}]^T$. The physical meaning of $U$ is that it transforms between the SK bases and the $[l^2,l_z]$ bases $[+2,+1,0,-1,-2]^T$, together with a local $\pi/2$-rotation on the $d$-orbital states with respect to $z$ axis. The whole symmetry operator $\tilde T$ includes an onsite spatial reflection along $x$ direction. The diagonal hopping matrix is given by Tablet III~\cite{SIHarrison}
\begin{table}[!hbp]
\begin{tabular}{|c|c|c|c|c|c|}
\hline
$(\sqrt{2})^5t^{(2)}_{\vec i\vec j}$(eV) & $d_{xy}$ & $d_{xz}$ & $d_{z^2}$ & $d_{yz}$ & $d_{x^2-y^2}$ \\
\hline
$d_{xy}$ & 0.2158 & 0 & 0 & -0.3603 & 0 \\
\hline
$d_{xz}$ & 0 & -0.5388 & 0.1138 & 0 & 0.1971 \\
\hline
$d_{z^2}$ & 0 & 0.1138 & 0.3630& 0 & -0.3689\\
\hline
 $d_{yz}$ & -0.3603 & 0 & 0 & 0.2158 & 0 \\
\hline
$d_{x^2-y^2}$ & 0 & 0.1971 & -0.3689 & 0 & -0.0629 \\
\hline
\end{tabular}
\caption{Diagonal hopping couplings for configuration (a)~\cite{SIHarrison}. Here $\vec j=\vec i+a\vec e_x+a\vec e_y$.}
\end{table}
From this tablet one can see that the diagonal hopping terms mix $d_{xy},d_{xz}$ states and other states, and such mixing explicitly breaks the aforementioned spatial reflection. One can verify that $\tilde T{\cal H}_{\rm triple}(t^{(2)}_{\alpha\alpha'},k_z)\tilde T^{-1}\neq{\cal H}_{\rm triple}(t^{(2)}_{\alpha\alpha'},-k_z)$ with $t^{(2)}_{\alpha\alpha'}\neq0$, and the $\tilde T$ symmetry is broken. The new BDI class is reduced to D class by the diagonal hopping couplings.

\subsection{Triple Fe-chain for the configuration (b)}

For the configuration (b) the spin is polarized along $\vec{y}$ direction, and the chains are also stacked along $y$ direction. In this configuration the magnetization is in-plane, and the diagonal hopping matrix is given in Tablet IV~\cite{SIHarrison}
\begin{table}[!hbp]
\begin{tabular}{|c|c|c|c|c|c|}
\hline
$(\sqrt{2})^5t^{(2)}_{\vec i\vec j}$(eV) & $d_{xy}$ & $d_{xz}$ & $d_{z^2}$ & $d_{yz}$ & $d_{x^2-y^2}$ \\
\hline
$d_{xy}$ & 0.2158 & 0.3603 & 0 & 0 & 0 \\
\hline
$d_{xz}$ & 0.3603& 0.2158 & 0 & 0 & 0 \\
\hline
$d_{z^2}$ & 0 & 0 & 0.3630& -0.1138 & 0.3689\\
\hline
 $d_{yz}$ & 0 & 0 & -0.1138 & -0.5388 & 0.1971 \\
\hline
$d_{x^2-y^2}$ & 0 & 0 & 0.3689 & 0.1971 & -0.0629 \\
\hline
\end{tabular}
\caption{Diagonal hopping couplings for configuration (b)~\cite{SIHarrison}. Here $\vec j=\vec i+a\vec e_x+a\vec e_y$.}
\end{table}
The $t^{(2)}_{ij}$ hopping terms are block diagonal and do not mix $d_{xy},d_{xz}$ states and other other states. Thus the diagonal hopping couplings cannot break the onsite spatial reflection along $x$ direction, and accordingly do not break the $\tilde T$ symmetry. On the other hand, the nearest- and next-next-nearest-neighbor couplings along $y$ direction take the forms in Tablets V and VI.
\begin{table}[t]
\begin{tabular}{|c|c|c|c|c|c|}
\hline
$t^{(1)}_{\vec i\vec j}$ (eV) & $d_{xy}$ & $d_{xz}$ & $d_{z^2}$ & $d_{yz}$ & $d_{x^2-y^2}$ \\
\hline
$d_{xy}$ & 0.5760 & 0 & 0 & 0 & 0 \\
\hline
$d_{xz}$ & 0& -0.1445 & 0 & 0 & 0 \\
\hline
$d_{z^2}$ & 0 & 0 & -0.2759& 0 & -0.2276\\
\hline
 $d_{yz}$ & 0 & 0 & 0 & 0.5760 & 0 \\
\hline
$d_{x^2-y^2}$ & 0 & 0 & -0.2276 & 0 & -0.5388 \\
\hline
\end{tabular}
\caption{Nearest-neighbor hopping coupling in $\vec{y}$ direction. Here $\vec j=\vec i+a\vec e_y$.}
\end{table}
\begin{table}[t]
\begin{tabular}{|c|c|c|c|c|c|}
\hline
$2^5t^{(3)}_{\vec i\vec j}$ (eV) & $d_{xy}$ & $d_{xz}$ & $d_{z^2}$ & $d_{yz}$ & $d_{x^2-y^2}$ \\
\hline
$d_{xy}$ & 0.5760 & 0 & 0 & 0 & 0 \\
\hline
$d_{xz}$ & 0& -0.1445 & 0 & 0 & 0 \\
\hline
$d_{z^2}$ & 0 & 0 & -0.2759& 0 & -0.2276\\
\hline
 $d_{yz}$ & 0 & 0 & 0 & 0.5760 & 0 \\
\hline
$d_{x^2-y^2}$ & 0 & 0 & -0.2276 & 0 & -0.5388 \\
\hline
\end{tabular}
\caption{Next-next-nearest-neighbor hopping coupling in $\vec{y}$ direction~\cite{SIHarrison}. Here $\vec j=\vec i+2a\vec e_y$.}
\end{table}
These terms also do not break the $\tilde T$ symmetry.

In the realistic system, a random magnetic disorder can exist with nonzero magnetization $\delta\vec J(\bold r)=\delta J^x\hat e_x$ along $x$ direction, which gives the disorder Hamiltonian
\begin{eqnarray}
V_{\rm dis}=\sum_{\alpha}\sum_{\bold r,\sigma\sigma'}c^\dag_{\alpha \sigma}(\bold r)\delta J^x(\bold r)s^x_{\sigma\sigma'}] c_{\alpha \sigma'}(\bold r).
\end{eqnarray}
This term transforms according to $\tilde TV_{\rm dis}\tilde T^{-1}=-V_{\rm dis}$. Therefore, the further inclusion of the magnetic disorder $V_{\rm dis}$ breaks the $\tilde T$ symmetry in the configuration (b). Note that the disorder with in-plane magnetization (along $y$ and $z$ directions) does not affect the symmetry.

\end{document}